\documentclass{WileyMSP-template}

\usepackage{amsmath} 
\usepackage{xcolor}
\definecolor{mark}{RGB}{205,222,255}

\begin{document}

\title{Long-term stability and oxidation of ferroelectric AlScN devices: An operando HAXPES study}

\maketitle

\author{Oliver Rehm}
\author{Lutz Baumgarten}
\author{Roberto Guido}
\author{Pia Maria Düring}
\author{Andrei Gloskovskii}
\author{Christoph Schlueter}
\author{Thomas Mikolajick}
\author{Uwe Schroeder}
\author{Martina Müller*}

\begin{affiliations}

Oliver Rehm, Pia Maria Düring, Prof. Dr. Martina Müller\\
Fachbereich Physik, Universität Konstanz, 78457 Konstanz, Germany\\
\,\\
Dr. Lutz Baumgarten\\
Forschungszentrum Jülich GmbH, Peter Grünberg Institut (PGI-6), 52425 Jülich, Germany \\
\,\\
Roberto Guido, Dr. Uwe Schroeder, Prof. Dr. Thomas Mikolajick  \\
NaMLab gGmbH, 01187 Dresden, Germany \\
\,\\
Dr. Andrei Gloskovskii, Dr. Christoph Schlueter \\
Deutsches Elektronen-Synchrotron, 22607 Hamburg, Germany \\
\,\\
Prof. Dr. Thomas Mikolajick \\
Technische Universität Dresden, 01062 Dresden, Germany \\

\end{affiliations}


\begin{abstract}

Aluminum scandium nitride (Al$_{1-x}$Sc$_x$N) is a promising material for ferroelectric devices due to its large remanent polarization, scalability, and compatibility with semiconductor technology. By doping AlN with Sc, the bonds in the polar AlN structure are weakened, which enables ferroelectric switching below the dielectric breakdown field. However, one disadvantage of Sc doping is that it increases the material's tendency towards oxidation.
In the present study, the oxidation process of tungsten-capped and uncapped Al$_{0.83}$Sc$_{0.17}$N thin films is investigated by hard X-ray photoelectron spectroscopy (HAXPES). The samples had been exposed to air for either two weeks or 6 months. HAXPES spectra indicate the replacement of nitrogen by oxygen, and the tendency of oxygen to favor oxidation with Sc rather than Al. The appearance of an N$_2$ spectral feature thus can be directly related to the oxidation process. We present an oxidation model that mimics these spectroscopic results of the element-specific oxidation processes within Al$_{1-x}$Sc$_x$N. Finally, in operando HAXPES data of uncapped and capped AlScN-capacitor stacks are interpreted using the proposed model.

\end{abstract}

\section{Introduction}
CMOS-compatible ferroelectric materials have only been discovered in the last ten years \cite{Boescke2011, Fichtner2019} which has spurred renewed interest and research into ferroelectric (FE) devices such as FE random access memories (FeRAM)\cite{Francois2021}, FE field effect transistors (FeFET)\cite{Mulaosmanovic2021} or FE tunnel junctions (FTJ)\cite{Ambriz-Vargas2017}. Scandium-doped AlN (AlScN) \cite{Fichtner2019} extends the growing field of nitride electronics \cite{Hickman2021, Haider2023} with FE functionality in terms of large polarization, large coercive fields and a nearly box-shaped hysteresis. Surprisingly, the natural polar state in AlN even has to be destabilized to make it switchable below the dielectric breakdown voltage. This can be achieved by lattice strain or doping, as has been first demonstrated using Sc by Fichtner et al. \cite{Fichtner2019}.

However, the switchable FE state that can be achieved in a device in this way suffers from a deterioration in performance degradation after relatively low switching rates \cite{Wang2022, Kim2023}. Indeed, although AlN is resistant to oxidation up to a temperature of around 800\,°C \cite{Lee2002}, AlScN shows considerable and rapid surface oxidation even at room temperature \cite{Ambriz-Vargas2017,Li2022,Wang2023}.
The modest stability against oxidation, which degrades the ferroelectric switching, turns out to be one of the biggest challenges for the application of AlScN devices \cite{Wang2023}. This applies not only to the protection of AlScN from air, but also the choice of electrode materials, such as metal oxides.

The surface oxidation of Al$_{1-x}$Sc$_x$N is well known and the process has been claimed to be self-limited \cite{Li2022} However, the influence of the oxidation process on the microscopic chemistry of AlScN has not yet been investigated in detail.
Fang et al. \cite{Fang2018} have theoretically investigated the oxidation of undoped AlN. In a proposed three-step model, a nitrogen ion is first removed from its lattice site, which is subsequently occupied by an oxygen atom. The released nitrogen then leaves the crystal as N${_2}$ molecule. 

In this study, we focused on the oxidation of AlScN as a particular disadvantage of the Sc doping of AlN. We investigated  Al$_{0.83}$Sc$_{0.17}$N samples that had been exposed to air for either weeks or months. Oxidation in air was chosen to mimic realistic device-fabrication conditions, during which the samples are not deposited in situ from the bottom to the top electrode but are exposed to air as they are transferred between growth chambers.

We used hard X-ray photoelectron spectroscopy (HAXPES) for the detailed investigation of the oxidation process of AlScN. Due to its element and chemical state selectivity together with the large information depth (ID), it allows a detailed investigation of the oxidation process \cite{Mueller2021}. The HAXPES experiments on samples exposed to air over different time periods are compared with reference samples of Al$_{0.83}$Sc$_{0.17}$N layers capped in-situ with a tungsten (W) film and an undoped AlN layer. We show a strong interdependence between the oxidation and the formation of nitrogen vacancies within the AlScN layer, which are suspected to play a decisive role on the field cycling reliability\cite{Wang2022, Kim2023}. These element-specific oxidation processes in Al$_{1-x}$Sc$_x$N are fed into a simple oxidation model that reproduces these spectroscopic results. However, a self-limiting oxidation process as proposed by Li et al.\cite{Li2022} we cannot confirm. In a final step, we make use of the proposed oxidation model to interpret first in operando HAXPES data of uncapped and capped AlScN-capacitor stacks.\\

\section{The N\,1s/Sc\,2p core level fingerprint}

To evaluate the long-term stability and oxidation behavior of Al$_x$Sc$_{1-x}$N in air, three samples were analysed. One sample was capped in-situ with W (3\,nm) and two additional samples were exposed to air for a fortnight and six months, respectively. The analysis focused on the Sc\,2p -- N\,1s and Al\,2s core levels, as they act as chemical fingerprints for the identification of oxidation-induced changes.

\begin{figure*}[htb]
\includegraphics[width=1\textwidth]{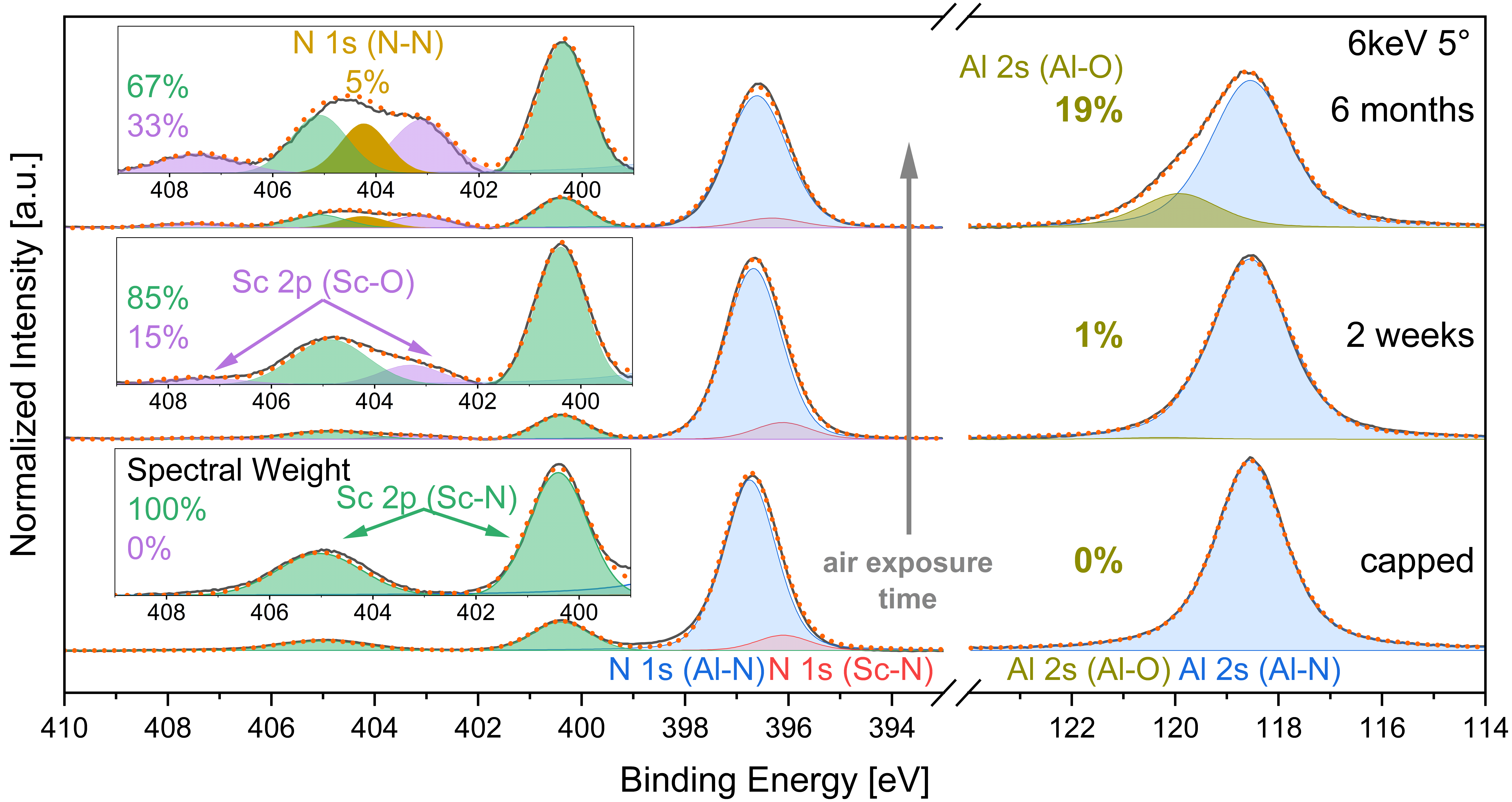}
\caption{HAXPES spectra of the Sc\,2p, N\,1s and Al\,2s core levels recorded at different stages of air exposure acquired 6\,keV photon energy and 5$^\circ$ electron emission angle. Peakfit results are underlayed. }
\label{fig:AlScN-Oxidation1}
\end{figure*}

\textbf{Figure \ref{fig:AlScN-Oxidation1}} shows the Sc\,2p, N\,1s and Al\,2s core levels for each sample at different stages of air exposure. The results demonstrate that air exposure led to remarkable changes in the peak shapes, especially in the Al\,2s core level at about 119\,eV, where an additional shoulder emerged, and in the Sc\,2p core level, where the changes become clearly apparent in the energy range from 399 to 409\,eV (see inset). 

The binding energy scale was calibrated using the Fermi edge of the W-capped Al$_{0.83}$Sc$_{0.17}$ sample. The uncapped samples were calibrated using the Al-N component of the Al\,2s emission, with the binding energies aligned accordingly. In order to align the uncapped samples, a shift towards lower binding energies was necessary. The observed shift in the sample that had been exposed to air for a period of two weeks was 790\,meV, while the observed shift in the sample that had been exposed for a period of six months was 540\,meV. Similarly, the same shift is applied to the Sc\,2p -- N\,1s region, which results in spectral alignment. Based on these observations, the previously described shift can be interpreted as a rigid binding energy shift resulting from a change in the valence band offset for AlScN \cite{Baumgarten2023}. 

The intensity is normalized to the integrated intensity of the Al\,2s emission in the case of Al\,2s and to the integrated intensity of the N\,1s - Sc\,2p emission, respectively. The N\,1s - Sc\,2p region is further scaled up by a factor of 1.3 relative to Al\,2s region.

The experimental spectra in figure \ref{fig:AlScN-Oxidation1} are overlayed with the results obtained from the peakfit analysis. The N\,1s emission should contain two contributions -- a Sc-N and a Al-N emission. This is confirmed by the peakfit results, since the N\,1s emission is slightly asymmetric and cannot be fitted by a single line. However, the chemical shift by replacing Al by a chemically similar Sc ion and the Sc-N intensity is expected to be small.  The intensity ratio of the underlying N\,1s peakfit contributions is estimated by the Al\,2s\,(Al-N)/Sc\,2p\,(Sc-N) intensity ratios, corrected by its spectral cross section. Comparing literature XPS values of AlN and AlScN \cite{Haseman2020,Haehnlein2020} the shift can be estimated to about 500\,meV. Our peakfit yields a chemical shift of about 550\,meV.

The Sc\,2p spin-orbit splitting and intensity ratio between $2p_{3/2}$ and $2p_{1/2}$, as obtained from the W-capped sample, are kept fixed for all spectral fits. In the case of the oxidized 2-week-old sample, a Sc-O doublet was considered in the fitting model in order to account for the oxidation. The line width slightly increases for the oxidized samples. The obtained spin-orbit splitting of Sc\,2p of 4.4\,eV, as well as the chemical shift of 2.8\,eV between Sc\,2p-Oxide (Sc-O) and Sc\,2p-Nitride (Sc-N), is in agreement with previously reported values \cite{Haseman2020,Haehnlein2020,Ding2023,Sana2016}.

The model of the 2-week-old sample was applied to the 6-month-old sample, and after 6 months of air exposure, the highly oxidized sample exhibited an additional emission at approximately 404\,eV binding energy in the Sc\,2p region. This peak needed to be accounted for because the fit parameter obtained from the W-capped and 2-week-old samples could not explain it. It is not associated with the Sc\,2p emission due to the absence of a spin-orbit doublet. The potential origin of this peak will be discussed below. It is also worth noting that the Sc-O share within the Sc\,2p core level spectra is larger than the Al-O share within the Al\,2s spectra.

To pinpoint the origin and development of the oxidation within the AlScN layer, we performed angle-dependent scans at a photon energy $h\nu = 6000$\,eV ($\Theta = 5{^\circ}$ and $30{^\circ}$) and at $h\nu = 2800$\,eV ($\Theta = 5{^\circ}$). This translates into information depths of 18\,nm, 15\,nm and 9\,nm, respectively.

\begin{figure*}[htb]
\includegraphics[width=1\textwidth]{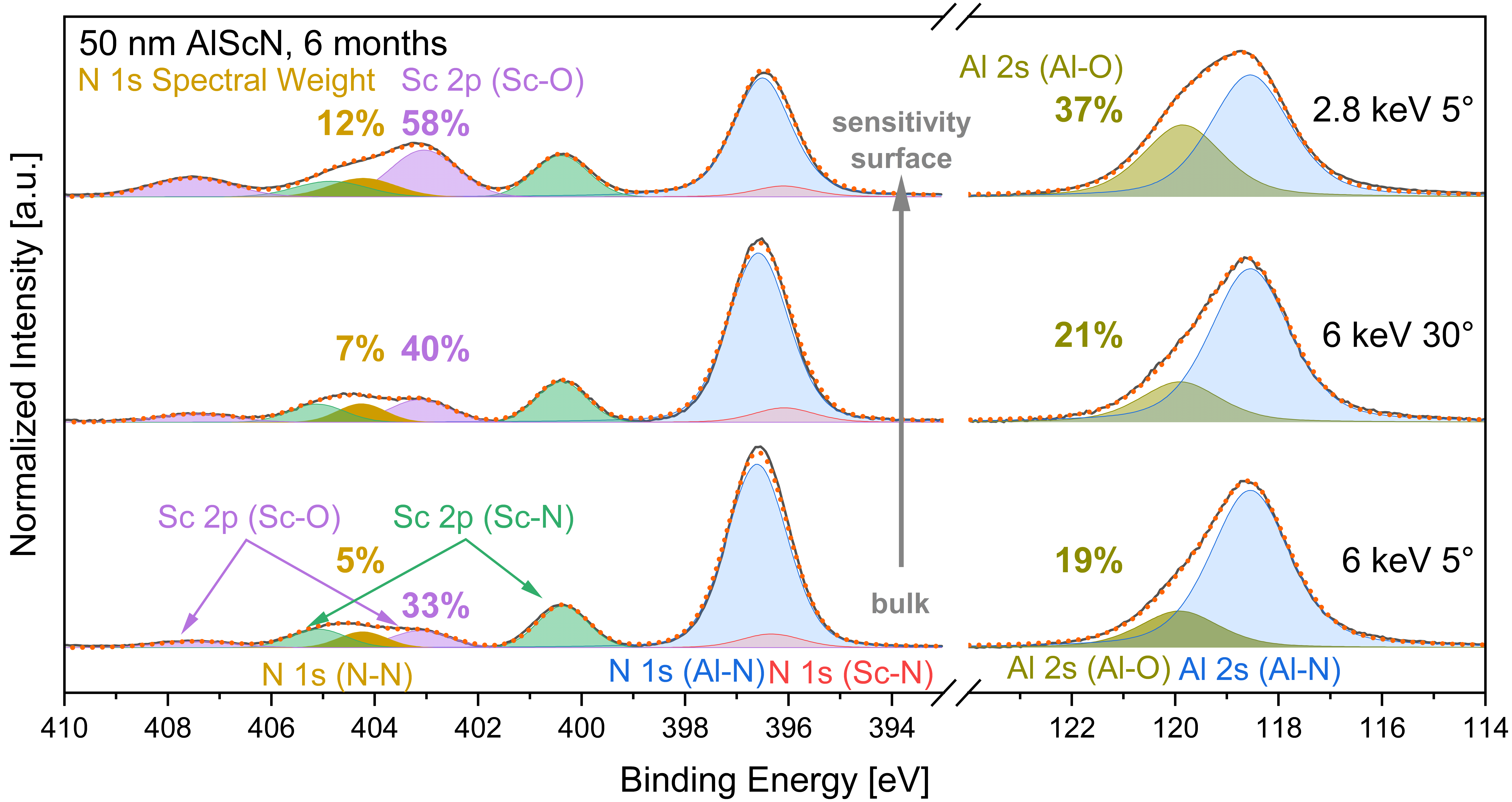}
\caption{Electron emission angle and photon energy dependent Sc\,2p, N\,1s and Al\,2s HAXPES spectra of a 6-month-old Al$_{0.83}$Sc$_{0.17}$N sample. The 6\,keV 5°, 6\,keV 30° and 2.8\,keV 5° spectra refer to an information depth of about 18\,nm, 15\,nm and 9\,nm, respectively.}
\label{fig:AlScN-Oxidation2}
\end{figure*}

\textbf{Figure \ref{fig:AlScN-Oxidation2}} illustrates the recorded HAXPES spectra of the Sc\,2p -- N\,1s and Al\,2s core levels in these experimental configurations for the 6-month-old sample. The Sc-O and Al-O spectral weight is enhanced for more grazing photoelectron emission and at $h\nu = 2800$\,eV, which corresponds to an enhanced surface sensitivity. The same behavior with a less oxide contribution is observed for the sample after 2 weeks of air exposure (not shown). Thus, obviously the intensity of the Sc-O and Al-O contribution continuously increases with air exposure time and is also enhanced at the surface.

This also holds true for the additional peak at 404.2\,eV, which is not directly related to the Sc\,2p emission but seems to be associated with the oxidation process. Notably, this peak is absent in the capped sample.

Furthermore, we investigated the effects of etching and oxidation on AlScN samples. Our findings revealed that the etching process, which could be performed employed during device fabrication of specialized devices, accelerates the oxidation rate. It is anticipated that the etching process increases the surface roughness of the sample, thereby creating an energetically more favorable environment for oxidation. Moreover, the etched sample yielded a signal from the bottom electrode. In the case of a Ti bottom electrode, it was found that the interface between AlScN and Ti is not stable due to the presence of a Ti-N signature in the Ti 2p spectra (not shown here). Further details on an analogue experimental analysis in HZO-based devices can be found here \cite{Vishnumurthy2024, Szyjka2020}.

\section{Origin of additional N\,1s Peak Feature}
The chemical origin of the additional peak in the Al$_{0.83}$Sc$_{0.17}$N 6-month-old sample has  been determined through a comparative analysis of the Al$_{0.83}$Sc$_{0.17}$N samples with a bare, undoped AlN sample that has been oxidized in air for six months.

\begin{figure}[htb]
\includegraphics[width=0.5\columnwidth]{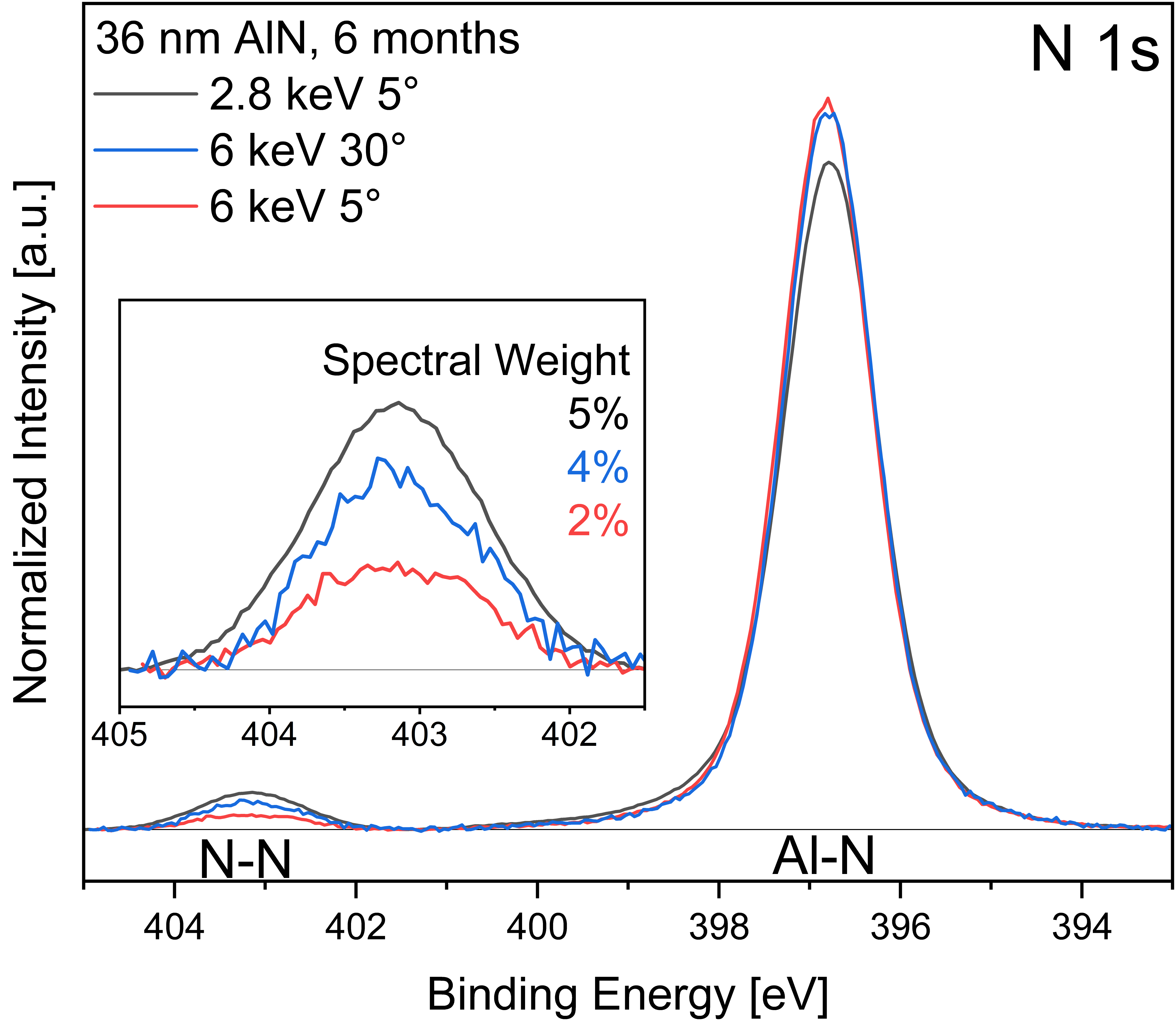}
\caption{N\,1s spectrum of AlN without Sc doping, taken at photon energies of 2.8\,keV and 6\,keV. The inset highlights the binding energy region of the N-N feature. The spectral weight refers to the total N\,1s intensity and increases with increasing surface sensitivity.}
\label{fig:AlN1}
\end{figure}

\textbf{Figure \ref{fig:AlN1}} depicts the N\,1s spectra of the bare, undoped AlN. In addition to the main N\,1s (Al-N) peak at 396.8\,eV, another peak is observed at 403.2\,eV. The splitting for AlN is 6.4\,eV, while the split for \\
Al$_{0.83}$Sc$_{0.17}$N was observed to be 7.7\,eV. This discrepancy can be explained by extra-atomic screening effects \cite{Esaka1998, Chung2005}. It can thus be concluded that the feature is part of the N\,1s core level and that is related to the oxidation of AlN and AlScN, respectively.

To summarise, the observations of the HAXPES spectra in figure \ref{fig:AlScN-Oxidation1}, \ref{fig:AlScN-Oxidation2} and \ref{fig:AlN1} are notable for two aspects in addition to the expected surface-enhanced oxidation. Firstly, it can be observed that the contribution of the Sc\,2p oxide is larger than that of the Al\,2s oxide. Secondly, the intensity of the additional N\,1s peak correlates with the oxide intensity and is not observed in the capped, non-oxidized sample. We will discuss the specific process of oxidation in Al$_{0.83}$Sc$_{0.17}$N below.

\section{Atomistic model for the oxidation of AlScN}

Understanding the oxidation behavior of Al$_{0.83}$Sc$_{0.17}$N is crucial for predicting its long-term stability and performance in various applications. In this section, we develop a basic model to describe the oxidation process by considering the differences in bonding strengths between Sc-N, Sc-O, Al-N, and Al-O  and taking into account the results from the HAXPES measurements. This model allows us to estimate the relative quantities of metal oxides and the amount of nitrogen released during oxidation. In addition, the proposed model provides insights into the depth of oxidation and the formation of oxide layers over time. In the following, we describe the theoretical framework and its implications, supported by experimental observations. Additionally, we also  discuss the limitations of the model.

The different oxidation ratios of Al and Sc can be explained by the different energy gain when replacing a Sc-N bond with a Sc-O bond (207.4 kJ\,mol$^{-1}$) compared to a replacement of Al-N by Al-O (133.9 kJ\,mol$^{-1}$) \cite{Luo2007}. 

\begin{figure*}[htb]
\includegraphics[width=0.9\textwidth]{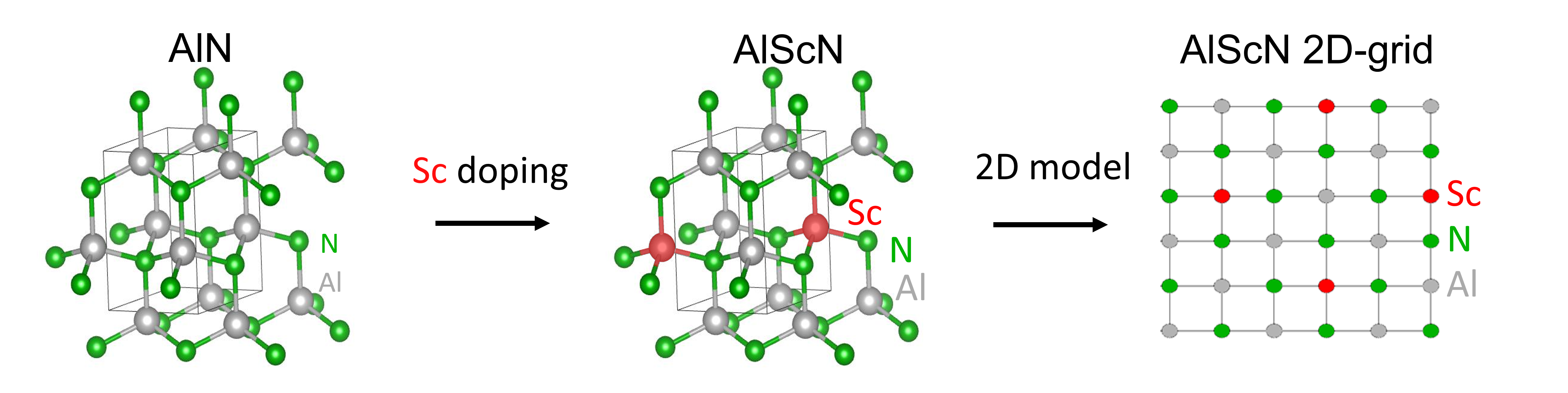}
\caption{Wurzit lattice structure of (a) AlN, (b) Al$_{0.83}$Sc$_{0.17}$N and (c) simplified 2D grid of Al$_{0.83}$Sc$_{0.17}$N.
The 2D grid reproduces the fourfold coordination of each metal and nitrogen atom.}
\label{fig:AlN2}
\end{figure*}

Now, we consider the oxidation process as a replacement of nitrogen ions by oxygen. In our basic oxidation model of AlScN, we assume that oxygen replaces only N-lattice sites next to Sc ions due to the larger energy gain. We now obtain the relative proportion of metal-oxides by counting the number of Sc-N, Sc-O, Al-N, and Al-O bonds. As depicted in \textbf{figure \ref{fig:AlN2}}, this is easily done in a two-dimensional AlScN lattice, where the fourfold coordination of metal and nitrogen ions is reproduced. Al ions are randomly replaced by Sc ions, under the assumption that no Sc ion occupies a lattice site next to another Sc ion, thus explicity excluding Sc-N-Sc bonds. The actual sample most likely contains Sc-N-Sc bonds.  We note that Sc-N-Sc bonds oxidize more easily due to the larger energy gain compared to Sc-N-Al and Al-N-Al.

The replacement of a nitrogen ion by oxygen thus only affects one Sc ion and results in one Sc-O bond and three Al-O bonds. For example, a 50\,\% Sc oxidation ($c_{ScO}$=0.5) of one Sc ion requires the replacement of two nitrogen ions by two oxygen ones, which in turn affects six Al-N bonds. At an Sc concentration of 20\,\% ($c_{Sc}$=0.2), the resulting Al-O concentration is expected to be 37.5\,\% ($c_{AlO}$=0.375). This can be easily formalized in general terms by
\begin{equation}
    \label{eq:concent1}
    c_{AlO} = 3 \cdot \frac{c_{Sc}\cdot c_{ScO}}{1-c_{Sc}}.
\end{equation}

As shown in \textbf{table \ref{tbl:oxidation}}, this basic model reproduces the reduced Al\,2s (Al-O) content quite well for the 6-month-old sample compared to the Sc\,2p (Sc-O) ratio.
Next, we consider the N\,1s (N-N) emission at about 404\,eV binding energy. Since each oxygen atom releases one nitrogen, the concentration of released nitrogen atoms ($c_N$) is given by
\begin{equation}
    \label{eq:concent2}
    c_{N} = 4 \cdot c_{Sc} \cdot c_{ScO}.
\end{equation}

The N-N concentration that is therefore expected is overestimated by a factor of about four. However, this expectation implies that the released N-N is completely stored in the lattice, which seems to be very unlikely. On the contrary, it is expected that most of the N$_2$ produced will be released during the oxidation process. Since the oxide layer is reportedly  very porous, this applies not only to the surface but also to deeper layers \cite{Yeh2017}. Therefore, an overestimation is to be expected by simply counting the released nitrogen atoms. Nevertheless, some of the gaseous nitrogen seemed to be stored as an interstitial site.

\begin{table*}[htb]
\centering
\caption{Comparison of observed and expected Al\,2s (Al-O) and N\,1s (N-N) element-specific spectral weights in percent depending on the observed Sc\,2p (Sc-O) spectral weight for Al$_{0.83}$Sc$_{0.17}$N with different air exposure times and information depths (ID). Calculation performed using \eqref{eq:concent1} and \eqref{eq:concent2}.}
\label{tbl:oxidation}
\begin{tabular}{|c|ccc|ccc|cc|}
\hline
\textbf{Oxidation time}  & \multicolumn{3}{c|}{\textbf{Measurement Parameters}}                                                                   & \multicolumn{3}{c|}{\textbf{Experiment}}                                                              & \multicolumn{2}{c|}{\textbf{Model}}                         \\
                        & \multicolumn{1}{c|}{\textbf{Energy [keV]}} & \multicolumn{1}{c|}{\textbf{Angle [°]}} & \textbf{ID [nm]} & \multicolumn{1}{c|}{\textbf{Sc-O [\%]}} & \multicolumn{1}{c|}{\textbf{Al-O [\%]}} & \textbf{N-N [\%]} & \multicolumn{1}{c|}{\textbf{Al-O [\%]}} & \textbf{N-N [\%]} \\ \hline
                        & \multicolumn{1}{c|}{2.8}                   & \multicolumn{1}{c|}{5}                  & 9                               & \multicolumn{1}{c|}{57.9}               & \multicolumn{1}{c|}{37.0}               & 11.7              & \multicolumn{1}{c|}{35.6}               & 39.4              \\ \cline{2-9} 
\textbf{6 months}       & \multicolumn{1}{c|}{6}                     & \multicolumn{1}{c|}{30}                 & 15                              & \multicolumn{1}{c|}{39.8}               & \multicolumn{1}{c|}{20.8}               & 6.6               & \multicolumn{1}{c|}{24.4}               & 27.1              \\ \cline{2-9} 
                        & \multicolumn{1}{c|}{6}                     & \multicolumn{1}{c|}{5}                  & 18                              & \multicolumn{1}{c|}{32.7}               & \multicolumn{1}{c|}{18.8}               & 5.4               & \multicolumn{1}{c|}{20.1}               & 22.2              \\ \hline
                        & \multicolumn{1}{c|}{2.8}                   & \multicolumn{1}{c|}{5}                  & 9                               & \multicolumn{1}{c|}{18.6}               & \multicolumn{1}{c|}{3.6}                & -                 & \multicolumn{1}{c|}{11.4}               & 12.7              \\ \cline{2-9} 
\textbf{2 weeks}        & \multicolumn{1}{c|}{6}                     & \multicolumn{1}{c|}{30}                 & 15                              & \multicolumn{1}{c|}{19.4}               & \multicolumn{1}{c|}{4.4}                & -                 & \multicolumn{1}{c|}{11.9}               & 13.2              \\ \cline{2-9} 
                        & \multicolumn{1}{c|}{6}                     & \multicolumn{1}{c|}{5}                  & 18                              & \multicolumn{1}{c|}{15.3}               & \multicolumn{1}{c|}{0.9}                & -                 & \multicolumn{1}{c|}{9.4}                & 10.4              \\ \hline
\end{tabular}
\end{table*}

In contrast, for the sample with two weeks of air exposure, neither the Al\,2s (Al-O) nor the N\,1s (N-N) spectral weights are well reproduced by the model. Here, the N\,1s (N-N) emission is below the detection limit, but an overestimation by the model is also to be expected. The overestimation of the Al-O concentration is an inherent deficiency of our simplified oxidation model. For a count of Al-O and Sc-O bonds, it is assumed that no Sc ions are adjacent. Therefore, no Sc-O-Sc bonds are formed. At a Sc concentration of about 17\,\%, this assumption can fail, and especially at an early stage of oxidation, Sc-O-Sc bonds can predominantly form due to the larger energy gain. This reduces the number of Al-O bonds that are formed by a single oxygen ion.  Nonetheless, the model reveals that Sc, especially within Sc-N-Sc bonding configurations, plays a crucial role in the early stages of oxidation.

The next important question concerns the depth of oxidation. Does the oxidation form a stable, self-limiting oxide overlayer or is it a continuous process that eventually leads to a completely oxidized AlScN layer? We can answer this question by comparing the measurements at different information depths at photon energies of 2.8\,keV and 6\,keV. If we assume a finite oxide overlayer, the thickness can be calculated from the signal attenuation of Al\,2s (Al-N) and Sc\,2p (Sc-N). To be consistent, a finite oxide overlayer should show the same thickness as calculated from the 2.8\,keV or 6\,keV measurement. Otherwise, both measurements will yield different thicknesses and thus falsify the assumption of a finite overlayer. We calculated the thickness of the hypothetical oxide overlayer for the 6-month-old sample as described by Szyjka et al. \cite{Szyjka2020} and found a calculated thickness of 1.2\,nm for the 2.8\,keV measurement and 1.5\,nm for 6\,keV. Thus, the assumption of a finite overlayer failed. From this, we conclude that there is an oxidation gradient and a continuous oxidation process that will eventually affect the entire AlScN layer and we cannot confirm a reported self-limiting process\cite{Li2022}.

\section{Oxidation during operando HAXPES}

\begin{figure*}[htb]
\includegraphics[width=1\textwidth]{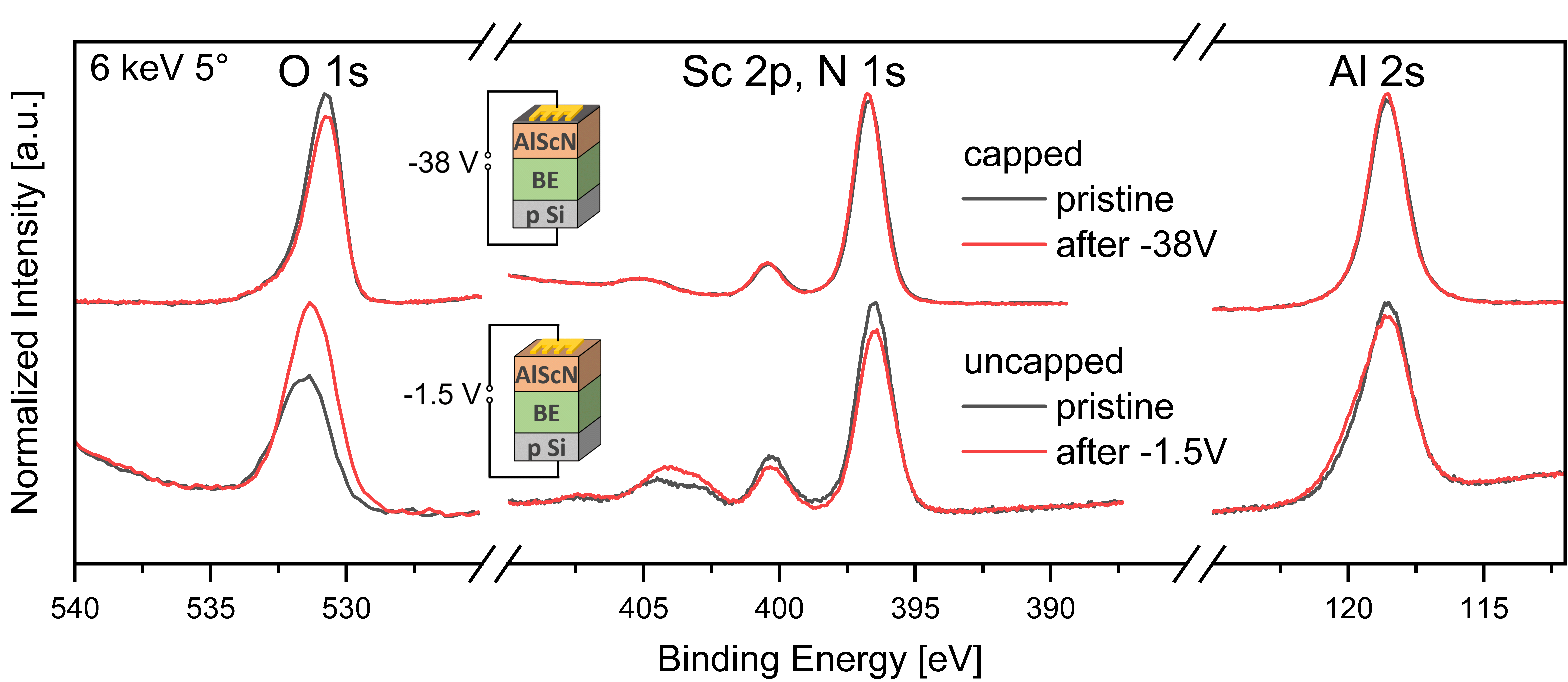}
\caption{Operando HAXPES measurements of an Al$_{0.83}$Sc$_{0.17}$N-based capacitor stack: Top panel shows capped (3\,nm W) AlScN with a maximum voltage of -38\,V applied, and bottom shows AlScN without a capping layer with a maximum voltage of -1.5\,V applied. The capped configuration demonstrates chemical stability under the applied voltage, whereas the uncapped, oxidized samples show an additional oxidation.}
\label{fig:operando1}
\end{figure*}

Oxidation of AlScN has a significant impact on device performance \cite{Liu2021,Guido2023}. Due to its non-self-limiting nature, continuous exposure to air will ultimately lead to the degradation of the entire device. Furthermore, the application of an electric field has the potential to influence the oxidation process significantly \cite{Liu2024}. Therefore, the selection of the top electrode, which effectively capped the AlScN layer is a crucial factor. If the top electrode on AlScN is damaged for instance, due to high-temperature exposure, any resulting crack would allow oxygen permeation and subsequent oxidation of the underlying AlScN \cite{Guido2023}. Consequently, electrode materials must be identified that are stable in contact with AlScN. This in particular applies to AlScN since, due to its high temperature stability, it is a promising candidate for high temperature applications.

Our operando experiment adresses the question wether and how an applied electric field influences the oxidation process. We compared a W capped, unoxidized layer with an oxidized sample. Both samples are equipped with a finger-shaped gold (Au) electrode to apply a homogeneous electric field over a large area (see sample schematics in figure \ref{fig:operando1}). The lateral electrode size is about 2\,mm by 2\,mm, while the x-ray beam spot is only 50\,µm by 150\,µm. The W-capping effectively prevents the AlScN layer for oxidation. Even after an air exposure of 6 month no oxide components within the Sc 2p core level are observable (see also figure \ref{fig:AlScN-Oxidation1}).

In \textbf{figure \ref{fig:operando1}}, the O\,1s, Sc\,2p - N\,1s, and Al\,2s core level regions of both an unoxidized sample capped by a 3\,nm W film and an oxidized AlScN sample are shown before and after in operando application of DC voltage. The uncapped, oxidized sample shows a significant increase in oxidation even upon application of only -1.5\,V, as indicated by the increased O\,1s, Sc\,2p (Sc-O) and Al\,2s (Al-O) contributions. Conversely, the unoxidized, capped sample remains stable up to about -38\,V and shows a reduction of the O\,1s core level signal. It should be noticed that we observe a large increase of the leakage current at -38\,V, which indicates the dielectric breakdown. Nevertheless, the chemical composition remains stable, which proves the stability of the W capping.

\section{Conclusion}
In summary, the long-term stability and oxidation process of tungsten-capped and uncapped Al$_{0.83}$Sc$_{0.17}$N thin films was investigated using HAXPES. Samples exposed to air for either two weeks or six months were investigated to mimic real ex-situ growth and device processing.

Our comprehensive HAXPES analysis of the Sc\,2p core level multiplet and the Al\,2s core level has provided the following insights into the oxidation process: Al$_{0.83}$Sc$_{0.17}$N is significantly sensitive to air, shows an oxidation gradient from the surface to the bulk, and undergoes accelerated oxidation as a result of Sc doping. 

An understanding of the oxidation mechanism could be achieved by considering the predominant substitution of nitrogen bound to scandium. The substituted nitrogen leads to the formation of N$_2$ molecules, which are mainly released from the sample. However, a small amount of N$_2$ remains in the lattice, presumably in an interstitial position. In general, we confirm the oxidation model as described by Fang et al. for undoped AlN \cite{Fang2018}. However, for AlScN the oxidation process is site-selective. Predominantly nitrogen bonded to Sc is released and its place is subsequently occupied by oxygen.

Applying an operando voltage to the partially oxidized sample during the HAXPES experiment leads to a considerable intensification of the oxidation process. 
In contrast, the AlScN layer, which protected by a 3\,nm tungsten capping layer, shows no signs of oxidation and remains chemically stable up to voltages of about 38\,V.

The results provide a comprehensive insight into the chemical oxidation processes in AlScN as an active layer in ferroelectric devices. The long-term stability of devices can be achieved through a complete in situ growth process and selecting appropriate materials for the electrodes. In this way, the ferroelectric functionality of AlScN is encapsulated and protected from the effect of air exposure, whereby the ferroelectric switching performance of devices can be improved beyond the current state of the art.

\section{Experimental Section}

The samples were prepared on p-Si substrates on which a 60nm thick bottom electrode (Ti, TiN, or W) was deposited by DC sputtering. Subsequently, Al$_x$Sc$_{1-x}$N films (60\,nm) were RF co-sputtered at 400\,°C using Al and Sc targets with a sputtering power of 200\,W and 140\,W, respectively. During the deposition the substrate rotated at 6\,rpm, and the nitrogen flow rate was 10\,sccm, which was twice as high as the argon flow rate of 5\,sccm.  
Consecutive deposition processes were performed without breaking the vacuum condition in an ultra-high vacuum sputter cluster tool from Bestec GmbH.

The stoichiometry was determined by HAXPES and calculated from the Al:Sc peak intensity ratio of the Sc\,2s, Sc\,3s, Sc\,3p, Al\,2s and Al\,2p peaks weighted by the respective photoionization cross sections \cite{Trzhaskovskaya_2018}. This procedure was performed using six combinations of different Al and Sc core levels and consistently yielded the stoichiometry Al$_{0.83}$Sc$_{0.17}$N for the uncapped Al$_x$Sc$_{1-x}$N sample with an accuracy of $\pm 4\,\%$.

HAXPES was performed at the P22 beamline of PETRA\,III (DESY, Hamburg)\cite{Schlueter2019} to investigate element-selective chemical properties. Core level spectra of Al, Sc, N, and O were recorded at a photon energy of 2.8\,keV and 6\,keV, providing an information depths  of 9\,nm and 18\,nm, respectively. The information depths were estimated using the Electron Spectra for Surface Analysis (SESSA) from the National Institute of Standards and Technology (NIST) \cite{Werner2017}.  A SPECS PHOIBOS 225HV electron analyzer was used at an emission angle of 5° and 30° and a pass energy of 50\,eV, resulting in an overall energy resolution of approximately 300\,meV. 

In operando HAXPES was conducted on specially prepared samples, on which a 'finger'-shaped gold (Au) top electrode (30\,nm) was deposited using a corresponding shadow mask. For the uncapped sample, an additional 4\,nm Au rectangle layer was deposited underneath to ensure a homogeneous voltage distribution across the area between the fingers.  This allowed the Al$_{0.83}$Sc$_{0.17}$N regions to be made accessible to the X-ray beam while voltage was applied on the electrode 'fingers'. The Au top electrode was connected to the sample holder electrode, while the p-Si was grounded on the holder. Voltage was applied using an Agilent B2912A power supply, with the DC voltage being gradually increased.

\medskip
\textbf{Conflict of Interest} \par
The authors declare no conflict of interest.

\medskip
\textbf{Acknowledgements} \par 
This work has also received funding from the BMBF (project 05K22VL1), by University of Konstanz BlueSky initiative, and by the VECTOR Foundation (project iOSMEMO). The authors acknowledge DESY (Hamburg, Germany), a member of the Helmholtz Association HGF, for the provision of experimental facilities. Beamtime was allocated for proposal(s) I-20230416 and I-20231120. Funding for the HAXPES instrument at beamline P22 by the Federal Ministry of Education and Research (BMBF) under contracts 05KS7UM1 and \\05K10UMA with Universität Mainz; 05KS7WW3, 05K10WW1, and 05K13WW1 with Universität Würzburg is gratefully acknowledged. Open access funding enabled and organized by Projekt DEAL. R.G. was financially supported by the Deutsche Forschungs Gemeinschaft within the project WUMM (project number: 458372836).

\medskip
\textbf{Data Availability Statement} \par 
The data that support the findings of this study are available from the corresponding author upon reasonable request. 

\medskip

\bibliographystyle{MSP}
\bibliography{AlScN.bib}

\end{document}